\documentclass[12pt]{article}
\usepackage[centertags]{amsmath}
\usepackage{amsfonts}
\usepackage{bm}
\begin{document}
\title{\text \bf{Local GL(d,R) Lie group for solving quantum ADM constraints}}
\author{H.S.Sharatchandra\thanks{E-mail:
{sharat@cpres.org}} \\[2mm]
{\em Centre for Promotion of Research,} \\
{\em 7, Shaktinagar Main Road, Porur, Chennai 600116, India}}
\date{}
\maketitle
\begin{abstract}
Relevance of local GL(d,R) Lie group for solving quantum ADM constraints is pointed out. Noncommutative harmonic analysis on $GL(d,R)/O(d)$ provides a natural basis and techniques for calculations. This is summarized. Such a basis is explicitly constructed for space dimension $d=2$.	
\end{abstract}

\smallskip

\smallskip


\section{Introduction}\label{i}
The Hamiltonian constraint of ADM formalism \cite{adm} in 3-space dimensions in absence of matter fields is 
\begin{eqnarray}
(q_{ac}(X)q_{bd}(X)p^{ab}(X)p^{cd}(X)-\frac{1}{2}(q_{bd}(X)p^{ab}(X))^2)-\frac{q(X)}{4\kappa^2}R^{(3)}(X)=0, \label{hc}
\end{eqnarray}
at each  point $X$ of 3-dimensional space. The momentum constraints are
\begin{eqnarray}
\partial_c(q_{ab}(X)p^{bc}(X))+\frac{1}{2}\partial_aq_{bc}(X)p^{bc}(X)=0. \label{mc}
\end{eqnarray}
Here $q_{ab}(X),p^{ab}(X), a,b=1,2,3$ are components of the 3-metric and its canonical conjugate respectively. 

In quantum gravity the metric and its conjugate variable are field operators $\hat q_{ab}(X),\hat p^{ab}(X)$ with canonical commutation relations 
\begin{eqnarray}
[\hat p^{ab}(X),\hat q_{cd}(Y)]=-i\hbar(\delta_{ac}\delta_{bd}+\delta_{ad}\delta_{bc})\delta^3 (X-Y),
\label{ccr}
\end{eqnarray}
with other commutators vanishing. The naive functional integral with Einstein action leads to quantum version \cite{I} of the constraints Eqn.~\ref{hc},\ref{mc}, with a symmetric ordering of $\hat q_{ab}$ and $\hat p^{ab}$. If we can 'solve' these constraints to get the 'physical states', then quantum gravity is dramatically simplified \cite{I}. In this paper we highlight the relevance of local GL(d,R) Lie group and noncommutative harmonic analysis on $GL(d,R)/O(d)$ for tackling the constraints.

Define formally a set of nine operator densities,
\begin{eqnarray}
\mathbf{T}_a^b(X)=\frac{1}{2\hbar}(\hat q_{ac}(X)\hat p^ {cb}(X)+\hat p^ {cb}(X)\hat q_{ac}(X)).
\label{t}
\end{eqnarray}
We have used a symmetric ordering to have formally self-adjoint operators. These composite operators formally have the algebra
\begin{eqnarray}
[\mathbf{T}_a^b(X),\mathbf{T}_c^d(Y)]= i(\delta_{ad}\mathbf{T}_c^b(X)-\delta_{bc}\mathbf{T}_a^d(X))\delta^3 (X-Y). \label{ga}
\end{eqnarray}
This is local $gl(3,R)$  Lie algebra, where $GL(3,R)$ is the general linear group on a 3-dimensional real vector space. The composite field $\mathbf{T}_a^b(X)$ can also be interpreted as the tensor $p_a^b(X)$ with mixed indices. But we do not want to do this.

The 'kinetic energy' part of the  Hamiltonian constraint Eqn.~\ref{hc} is remarkable in many ways. i. It is ultralocal. In any kind of discretization it can be treated independently at each $X$ and then the continuum limit can be taken. ii. It is scale invariant. iii. It appears to be related to Casimirs $\mathbf{T}_a^b(X)\mathbf{T}_b^a(X)$ and $\mathbf{T}_a^a(X)$. Therefore we first address (in Sec.\ref{g}) zero space dimensional version of this part. In Sec.\ref{2} we obtain explicit basis for $GL(2,R)$. In Sec.\ref{ha} we show the connection to noncommutative harmonic analysis on $GL(d,R)/O(d)$ and give a fast summary of its techniques and results. Finally a short discussion of the relevance of these techniques for solving solving quantum ADM constraints is given in Sec.\ref{dis}. 

\section{$gl(d,R)$ Lie algebra}\label{g}
Consider a set of $d$ quantum mechanical position and momentum operators labelled  $\hat q_{ab},\hat p^{ab}, a,b=1,2,\cdots d$, with $\hat q_{ab}=\hat q_{ba},\hat p^{ab}=\hat p^{ba}$,
\begin{eqnarray}
[\hat p^ {ab},\hat q_{cd}]=-i\hbar (\delta_{ac}\delta_{bd}+\delta_{ad}\delta_{bc}).\label{c}
\end{eqnarray}
With this definition there is an extra  factor of $2$ on rhs in the canonical commutaton rules for diagonal elements like $p^ {11},q_{11}$. Define $d^2$ self-adjoint operators 
\begin{eqnarray}
\mathbf{T}_a^b=\frac{1}{2\hbar}(\hat q_{ac}\hat p^ {cb}+\hat p^ {cb}\hat q_{ac})=\frac{1}{\hbar}\hat q_{ac}\hat p^ {cb}- i\frac{d+1}{2}\delta_{ab}. \label{t}
\end{eqnarray}
These have the algebra
\begin{eqnarray}
[\mathbf{T}_a^b,\mathbf{T}_c^d]= i(\delta_{ad}\mathbf{T}_c^b-\delta_{bc}\mathbf{T}_a^d). \label{l}
\end{eqnarray}
This is the Lie algebra $gl(d,R)$.

The transformation properties of $\hat q_{ab},\hat p^{ab}$ are
\begin{eqnarray} 
[\mathbf{T}_a^b,\hat q_{cd}]=-i(\delta_{bc}\hat q_{ad}+\delta_{bd}\hat q_{ca}), ~~
[\mathbf{T}_a^b,\hat p^{cd}]=i(\delta_{ac}\hat p^{bd}+\delta_{ad}\hat p^{cb}). \label{ttr}
\end{eqnarray}
This means the following: under the action of an element $G(\xi)=exp(i\mathbf{T}_a^b \xi_b^a)$ of $GL(d,R)$,
\begin{eqnarray}\label{gtq}
G\hat q_{ab}G^{-1}=(g\hat q\tilde g)_{ab},~ 
G\hat p^{ab}G^{-1}=((\tilde g)^{-1}\hat pg^{-1})^{ab}.\label{gtr} 
\end{eqnarray}
Here $g(\xi)=exp(it_a^b \xi_b^a)$, with $(t_a^b)$ as the generators of $GL(d,R)$ in the defining representation, $(t_a^b)_c^d=-i\delta_{ad}\delta_{bc}$.
$\tilde g$ is the transpose of matrix $g$.  Using the terminology of general relativity,  we say that $q_{ab}$ transforms covariantly  and $p^{ab}$
contravariantly as  symmetric tensors under $GL(d,R)$.

Now consider transformation property of $q^{ab}$ (elements of the matrix inverse of $q_{ab}$,) under $GL(d,R)$. Since for an infinitesimal variation, 
$\delta q^{ab}=-q^{ac}\delta q_{cd}q^{db}$,
using Eqn.~\ref{ttr} we get $[\mathbf{T}_a^b,\hat q^{cd}]=i(\delta_{bc}q^{ad}+\delta_{bd}\hat q^{ca})$,
which is the contravariant transformation. Thus the subscripts correspond to a covariant  transformation and the superscripts, a contravariant transformation, as in case of   general relativity.
It is relevant to consider the transformation property of $q=det(q_{ab})$. Using
$\delta q^r=r q^r q^{ab}\delta q_{ba}$,
where $r$ is an arbitrary  real or complex number we get,
\begin{eqnarray} 
[\mathbf{T}_a^b,q^r]=-2ir q^r\delta_{ab}. 
\label{tq}
\end{eqnarray}
The trace element 
\begin{eqnarray} 
\mathbb T=\sum_a \mathbf{T}_a^a,
\label{bbt}
\end{eqnarray}
commutes with all the generators. The traceless combinations,
\begin{eqnarray} 
{\mathcal T}_a^b=\mathbf{T}_a^b-\frac{1}{d}\delta_{ab}{\mathbb T},
\label{calt}
\end{eqnarray}
have the Lie algebra $sl(d,R)$ of the special linear group. Now,
\begin{eqnarray}
[\mathcal{T}_a^b,\hat q^r]=0;~ 
[\mathbb{T},\hat q_{ab}]=-2i\hat q_{ab};~
[\mathbb{T},\hat q^r]=-2dri\hat q^r. 
\label{bbcal}
\end{eqnarray}
This means i. $q$ behaves like a c-number under $SL(d,R)$ transformations; ii. $q_{ab}$ scales with a weight $-2$ under $\mathbb{T}$ and iii. $q^{r}$ scales with a weight $-2dr$ under $\mathbb{T}$. Note that any diagonal generator 
$\mathbf{T}_a^a, ~a=1,2,\cdots,d$, counts the number of $a$ super- (sub-) scripts with a weight 1 (-1). Therefore $\mathbf{T}$ counts the number of {\it any} super- (sub-) script with a weight 1 (-1).

We now consider 'wave functions' $\psi(\{q_{ab}\})$ which have nice transformation properties under $GL(d,R)$. First we construct the singlet representation $\psi_0(q)$ which is invariant under $GL(d,R)$. As the self-adjoint choice of $\mathbf{T}_a^b$ is given by Eqn.~\ref{t}, we need to solve the equation
\begin{eqnarray} 
(\hat q_{ac}\hat p^ {cb}-  i\frac{d+1}{2}\delta_{ab})\psi_0(q)=0.
\label{epsi0}
\end{eqnarray}
We have $\hat p^ {cb}q^r=-2ir\delta_{cb}q^r$,
so that
\begin{eqnarray} 
\mathbf{T}_a^b q^r =-2i(r +\frac{d+1}{4})\delta_{ab}q^r. 
\label{tq}
\end{eqnarray}
Thus the singlet representation is given by the wave function  
\begin{eqnarray}
\psi_0(q)=q^{-(d+1)/4}.  \label{psi0}
\end{eqnarray}
This is annihilated by all generators.

Consider simplest case $d=1$ of one quantum mechanical position and momentum operator to see the novel issues that appear in case of non-compact groups. 
\begin{eqnarray}
\hat t=\frac{1}{2\hbar}(\hat q\hat p+ \hat p\hat q) \label{t1}
\end{eqnarray}
is the self-adjoint dilatation operator and $\hat q$ has a scale dimension $-1$. Now the wave function $\phi_0(q)$ annihilated by $\hat t$ satisfies
\begin{eqnarray} 
-i(q\frac{d}{dq}+\frac{1}{2})\phi_0(q)=0.
\label{phi0}
\end{eqnarray}
We get $\phi_0(q)\sim 1/\sqrt{q}$.
The complete set of eigenstates of $\hat t$ are found by solving the differential equation
\begin{eqnarray} 
-i(q\frac{d}{dq}+\frac{1}{2})\phi_r(q)=r\phi_r(q),
\end{eqnarray}
and therefore
\begin{eqnarray} 
\phi_r(q)=\frac{1}{\sqrt{2\pi q}}q^{ir}.\label{1b}
\end{eqnarray}
$r$ can have any complex value. As $\hat t$ is Hermitian, we allow only real eigenvalues, $r\in(-\infty,\infty)$. We also notice that the natural variable is $t=ln\, |q|$, with the  dilatation group acting as a translation of $t$. Also $q\in (0,\infty) $ and $q\in (-\infty,0)$ are separately invariant under the group action. Consider the range $q \in (0,\infty)$. Now $t$ is real and has the range $(-\infty,\infty)$. With all this the basis Eqn.~\ref{1b} is complete,
\begin{eqnarray} 
\int_{0}^{\infty} dq ~\phi_r(q)^*\phi_s(q)  =\int_{-\infty}^{\infty} \frac{dt}{2\pi} e^{-i(r-s)t} = \delta(r-s).
\label{on}
\end{eqnarray}
The message is the following: The singlet representation $\phi_0(q)$ serves to set the correct measure $1/\sqrt{2\pi q}$. 
The Hibert space relevant for the action of the group is the space of square integrable functions of $ln ~q$. The basis functions $\phi_r(q)$ (in particular the singlet representation) are not square integrable, in analogy with the eigenstates of position or momentum operator in quantum mechanics.

These considerations are directly relevant to the centre $GL(1,R)$ part of $GL(d,R)$ in our case.
\begin{eqnarray} 
\mathbb{T}=\frac{1}{2\hbar}\sum_{ab}(\hat q_{ab}\hat p^ {ab}+\hat p^ {ab}\hat q_{ab}).
\label{bbd}
\end{eqnarray}
This is a self-adjoint dilatation operator, with scale dimension $-2$ for the diagonal variables $\hat q_{aa}$ (due to the commutation relations Eqn.~\ref{ccr}) and scale dimension $-1$ for the off-diagonal variables $\hat q_{ab}, b\neq a$. Eqn. \ref{bbcal} means that $q=det(q_{ab})$ acts as a scalar wrt $SL(d,R)$ subgroup and can be used to construct representations of $GL(1,R)$. The trivial representation is given by Eqn.~\ref{psi0}. Also
\begin{eqnarray} 
\mathbb{T}\psi_r (q)=r\psi_r (q),~
\psi_r(q)= q^{-\frac{d+1}{4}+i\frac{r}{2d}}.
\label{bbd}
\end{eqnarray}
We are only interested in positive definite metrics, $q>0$. Therefore the range of $q$ is naturally $(0,\infty)$.

For a infinitesimal symmetric matrix $\delta q_{ab}$ the $GL(d,R)$ invariant metric is
\begin{eqnarray}
<\delta q, \delta q>=tr(\delta q q^{-1}\delta qq^{-1})=\sum_{abcd}\delta q_{ab}q^{bc} \, \delta q_{cd}q^{da}.
\label{im}
\end{eqnarray}
This gives $GL(d,R)$ invariant functional measure
\begin{eqnarray}
D q= \prod_{ab}dq_{ab}~|det M|^{1/2},
\label{im}
\end{eqnarray}
where $M$ is the $\frac{d(d+1)}{2}\times \frac{d(d+1)}{2}$ symmetric matrix
\begin{eqnarray}
M^{ab,cd}=\frac{1}{2}(q^{ac} q^{bd}+q^{ad} q^{bc}).
\label{M}
\end{eqnarray}
Now dimension counting gives $|det M|\propto q^{-(d+1)}$. Thus the $GL(d,R)$ invariant measure is 
\begin{eqnarray}\label{m1}
D q= q^{-(d+1)/2}\prod_{ab}~dq_{ab},
\label{md}
\end{eqnarray}
and the inner product for functions of $\{q_{cd}\}$ is,
\begin{eqnarray}
<\phi, \psi>=\int q^{-(d+1)/2}\prod_{ab}~dq_{ab}~ \phi^*(\{q_{cd}\}) \psi(\{q_{cd}\}).
\label{ipd}
\end{eqnarray}

\section{$d=2$: Explicit basis}\label{2}
We now obtain a complete basis at a point of space in 2-space dimensions, for which the relevant group is $GL(2,R)$.

We have $SL(2,R)$ as the invariance group of $det \{q_{ab}\}$ under the transformation $q_{ab}\rightarrow (gq\tilde g)_{ab}$.
Choose new variables
\begin{eqnarray} 
U=\frac{1}{2}(q_{11}+q_{22}),~V=\frac{1}{ 2}(q_{11}-q_{22}),~W=q_{12}.
\label{uvw}
\end{eqnarray}
We have
\begin{eqnarray} 
q=U^2-V^2-W^2.
\label{uvwi}
\end{eqnarray}
This presents the action of $SL(2,R)$ as the Lorentz group $SO(2,1)$ keeping the 'metric' $q$ invariant. Define scaled variables
\begin{eqnarray} 
u=\frac{U}{\sqrt q}, v=\frac{V}{\sqrt q},w=\frac{W}{\sqrt q}.
\label{uvws}
\end{eqnarray}
We have self-adjoint generators of
boost in u-v plane, boost in u-w plane and rotation in v-w plane respectively as
\begin{eqnarray} 
L_1=-i(u\frac{\partial}{\partial v}+v\frac{\partial}{\partial u}),~
L_2=-i(u\frac{\partial}{\partial w}+w\frac{\partial}{\partial u}),~
L_0=-i(v\frac{\partial}{\partial w}-w\frac{\partial}{\partial v}),
\label{l}
\end{eqnarray}
with the commutators
\begin{eqnarray} 
[L_0,L_1]=iL_2,~[L_0,L_2]=-iL_1,~[L_1,L_2]=-iL_0.\label{lc}
\end{eqnarray}
The opposite sign in the last commutator differentiates the non-compact $SO(2,1)$ Lie algebra from the compact case of angular momentum algebra $SO(3)$.

Using Eqn. \ref{uvw} we can relate these to our generators Eqn. \ref{t}: 
\begin{eqnarray} 
L_1=\frac{1}{2}(\mathcal{T}_1^1-\mathcal{T}_2^2)=\mathcal{T}_1^1=-\mathcal{T}_2^2,~L_2=\frac{1}{2}(\mathcal{T}_1^2+\mathcal{T}_2^1),~L_0=\frac{1}{2}(\mathcal{T}_1^2-\mathcal{T}_2^1).\label{lt}
\end{eqnarray}
The $SO(2,1)$ invariant measure is
\begin{eqnarray} 
\int dU~ dV~ dW~  \delta(U^2-V^2-W^2-q)=q^{1/2}\int du~ dv~ dw~  \delta(u^2-v^2-w^2-1).\label{imso}
\end{eqnarray}
Our generators are self-adjoint with this inner product.

Choose the parametrization,
\begin{eqnarray} 
u=cosh ~\chi, ~v=sinh ~\chi ~cos ~\theta, ~w=sinh ~\chi ~sin ~\theta.\label{uvwp}
\end{eqnarray}
We get
\begin{eqnarray} 
L_1=-i(cos ~\theta ~\frac{\partial}{\partial \chi}-coth ~\chi ~sin ~\theta ~\frac{\partial}{\partial \theta }),~ 
L_2=-i(sin ~\theta ~\frac{\partial}{\partial \chi}+coth ~\chi ~cos ~\theta ~\frac{\partial}{\partial \theta }),~ 
L_0=-i~\frac{\partial}{\partial \theta}.\label{lp}
\end{eqnarray}
The Casimir operator is $C_2=\frac{1}{2}\mathcal{T}_a^b\mathcal{T}_b^a=L_1^2+L_2^2-L_0^2 $. As with  the angular momentum algebra, we consider the common eigenvalues of $C_2,L_0$.
Choose the ansatz
$\psi (\theta, \chi)=e^{im\theta}f(\chi)$. For single-valuedness we have only integer values for $m$, $m=0, \pm 1,\pm 2, \pm 3, \cdots$. It is the eigenvalue of $L_0$. The eigenvalue equation $C_2 \psi= \lambda\psi$ gives
\begin{eqnarray} 
-\frac{d^2f}{d \chi^2}-coth \chi ~\frac{df}{d \chi}-m^2 sech^2\chi ~f=\lambda f.\label{eqf}
\end{eqnarray}
This ordinary differential equation can be rewritten as
\begin{eqnarray} \label{l}
(1-u^2)\frac{d^2f}{d u^2}-2u\frac{df}{du }+(\nu (\nu+1)-\frac{m^2}{1-u^2})f=0,\label{equ}
\end{eqnarray}
where $\lambda =-\nu (\nu+1)$. This is the Legendre equation \cite{nist} with two linearly independent solutions $\it P_{\nu}^{m}(u), \it Q_{\nu}^{m}(u)$.
These are the associated Legendre functions of the first and second kind, of degree $\nu$ and order $m$ respectively. (Our range of $u$ is $(1, \infty)$, and therefore we need associated Legendre functions and not the Ferrer's functions.) We have
\begin{eqnarray}
\nu=-\frac{1}{2}\pm\sqrt{\frac{1}{4}-\lambda}\label{nu}.
\end{eqnarray}
Legendre equation is invariant under $\nu+\frac{1}{2} \leftrightarrow -(\nu+\frac{1}{2})$. Thus it is sufficient to choose only one of the signs above. $C_{2}$ is self-adjoint
but not positive definite. Therefore $\lambda$ is real but not required to be non-negetive. This means the possible ranges for $\nu$ are
\begin{eqnarray}
Case~ 1.~\nu=-\frac{1}{2}+is;~s\in (0,\infty) ~if~
\lambda \in (\frac{1}{4},\infty). \\\nonumber
Case~ 2.~\nu \in(-\frac{1}{2},\infty)~ if ~
\lambda \in (\frac{1}{4},-\infty). \label{nu1}
\end{eqnarray}
Case~ 1. $\nu=-\frac{1}{2}+is$:~${\it P}_{-\frac{1}{2}+is}^{m}(u)$ is real valued and known as Conical or Mehler function \cite{nist}. Generalized Mehler-Fock transform is defined as
\begin{eqnarray}   
F(s)=\frac{s}{\pi}sinh (s\pi)~ \Gamma(\frac{1}{2}-m+is)\int_1^{\infty}{\it P}_{-\frac{1}{2}+is}^{m}(u) f(u) du.\label{mf}
\end{eqnarray}
for functions $f(u)$ such that $f(u)~ln(1+u) /\sqrt u \in L^1(1,\infty)$. For $m= \pm 1/2$ this corresponds to the Fourier cosine and sine transforms respectively. Therefore it is a generalization of Fourier transform. We can recover the function $f(u)$ by the inversion formula
\begin{eqnarray} 
f(u)=\int_0^{\infty}{\it P}_{-\frac{1}{2}+is}^{m}(u) F(s) ds.\label{mfi}
\end{eqnarray}
The $SO(2,1)$ invariant measure is
\begin{eqnarray} 
q^{1/2}\int du~ d(v^2+w^2)~ d\theta~  \delta(u^2-(v^2+w^2)-1)=\frac{1}{q^{3/2}}\int_1^{\infty} du~\int_0^{2\pi}d\theta.~\label{b2}
\end{eqnarray}
Case 2. $\nu \in(-\frac{1}{2},\infty)$: The Legendre function that does not grow at infinity is,
\begin{eqnarray}
{\it Q}_{\nu}^{m}(u)\sim u^{-\nu-1}.\label{qinf}
\end{eqnarray}
We also need to consider the behaviour as $u \rightarrow 1+$. 
\begin{eqnarray} 
{\it Q}_{\nu}^{m} \sim (u-1)^{-m/2}.\label{q1}
\end{eqnarray}
Define
\begin{eqnarray}
L_\pm=L_1\pm iL_2=-ie^{\pm i\theta} (\frac{\partial}{\partial \chi}\mp im~coth ~\chi). \label{lpm}
\end{eqnarray}
\begin{eqnarray}
[L_0,L_\pm]=\pm L_\pm, [L_+,L_-]=-2L_0.\label{lpmc}
\end{eqnarray}
Thus the generators $L_\pm$ raise and lower the eigenvalue $m$ respectively by $1$. We need a function space on which these generators repeatedly act. Because of the bad singularity Eqn. \ref{q1} as $u \rightarrow 1+$, Case 2 is ruled out.

We now relate this to the classification \cite{h1,h2,d} of all irreducible unitary representations of $SL(2,R)$. We have the following types.
\begin{eqnarray} 
I. Continuous ~series:
i. ~Principal ~series:\lambda\ge \frac{1}{4}, \nu=-\frac{1}{2}+is, \\ \nonumber
ii. ~Complementary ~series:\lambda\le \frac{1}{4}, -\frac{1}{2} \le \nu \le 0,\\ \nonumber
II. ~Discrete ~series:\nu \ge 0, m=\pm \nu, \nonumber\label{ui}
\end{eqnarray}
apart from the trivial representation. Realization of the group on the hyperbolic space $H^2$ has only $(I.i)$. See Sec.\ref{ha} for further discussion.

We now collect these results to obtain a complete basis in the case of $GL(2,R)$ in which $\mathbb T,L_0, tr(\mathcal{T}_a^b\mathcal{T}_b^a)$ are diagonalized. 
\begin{eqnarray}
\psi_{r,s,m}({q_{ab}})=q^{-3/4+ir/4}(\frac{q_{11}-q_{22}-2iq_{12}}{q_{11}-q_{22}+2iq_{12}})^{im}{\it P}_{-\frac{1}{2}+is}^{m}(\frac{q_{11}+q_{22}}{2\sqrt q})\label{ipd}
\end{eqnarray}
We represent this state, appropriately normalized, by the ket vector $|r,s,m>$.
\begin{eqnarray} 
\mathbb{T}|r,s,m>=r|r,s,m>,
~L_0|r,s,m>=~m|r,s,m>,
~C_2|r,s,m>=~(\frac{1}{4}+s^2)|r,s,m>\label{ipd}
\end{eqnarray}
We have the action of raising and lowering operators,
\begin{eqnarray} 
L_\pm|r,s,m>=\sqrt{((m\pm\frac{1}{2})^2+s^2)}~|r,s,m\pm 1>
\end{eqnarray}

Even though this is a perfectly viable and simple basis, it is not useful for addressing the ADM constraints. The reason is we need these labels at each point of space $X$. As $m$ is a discrete (integer) label, we do not get a smooth and continuous label. Therefore we use a different basis, where $(\mathbf{T}_1^1-\mathbf{T}_2^2)/2$ is diagonalized instead of $L_0$. $L_0$ is the generator of the compact $SO(2)$ subgroup of $SL(2,R)$ and therefore its spectrum labelled by $m$ were discrete. Instead, $\frac{1}{2}(\mathbf{T}_1^1-\mathbf{T}_2^2)$ is the generator of the non-compact $SL(1,R)$ subgroup, and its spectrum would be a continuous label. Indeed it is a scaling operator, which scales $q_{11}$, $q_{12}$ and $q_{22}$ with weights $-1,0,1$ respectively. $L_0$ and $(\mathbf{T}_1^1-\mathbf{T}_2^2)/2$  are generators of distict Cartan subgroups which are not related by conjugation. We have, 
\begin{eqnarray} 
\mathcal T={\mathcal T}_1^1=-{\mathcal T}_2^2=\frac{1}{2}(\mathbf{T}_1^1-\mathbf{T}_2^2),
~{\mathcal T}_1^2=\mathbf{T}_1^2, ~{\mathcal T}_2^1=\mathbf{T}_2^1.\label{ipd}
\end{eqnarray}
The transformation of $q_{ab}$ under these generators is, 
\begin{eqnarray} 
[\mathcal T,q_{11}]=-iq_{11},
[\mathcal T,q_{22}]=iq_{22},
[\mathcal T,q_{12}]=0,\\ \nonumber
[\mathcal T_1^2,q_{11}]=0,
[\mathcal T_1^2,q_{22}]=-2iq_{12},
[\mathcal T_1^2,q_{12}]=-iq_{11},\\ \nonumber
[\mathcal T_2^1,q_{11}]=-2iq_{12},
[\mathcal T_2^1,q_{22}]=0,
[\mathcal T_2^1,q_{12}]=-iq_{22}.\label{ipd}
\end{eqnarray}
 We use the Ansatz
\begin{eqnarray}
\psi(\{q_{ab}\})=q^{-3/4+ir/3}(\frac{q_{11}}{q_{22}})^{it/2}f(\frac{q_{12}}{\sqrt q}).\label{ipd}
\end{eqnarray}
$\mathcal T$ counts the difference in the number of $1$ and $2$ subscripts: $\mathcal T\psi=t\psi$.
We get 
\begin{eqnarray}
\mathcal T_1^2\psi(\{q_{ab}\})=q^{-3/4+ir/3}(\frac{q_{11}}{q_{22}})^{it/2}(-i\frac{q_{11}}{\sqrt q}f'-t\frac{q_{12}}{q_{22}}f).\label{ipd}
\end{eqnarray}
\begin{eqnarray}
\mathcal T_2^1\psi(\{q_{ab}\})=q^{-3/4+ir/3}(\frac{q_{11}}{q_{22}})^{it/2}(-i\frac{q_{22}}{\sqrt q}f'
+t\frac{q_{12}}{q_{11}}f).\label{ipd}
\end{eqnarray}
The Casimir is,
\begin{eqnarray} 
C_2=\frac{1}{2}\mathcal T_a^b\mathcal T_b^a=\mathcal T_1^2\mathcal T_2^1+\mathcal T^2+i\mathcal T.
\end{eqnarray}
We get,
\begin{eqnarray}
C_2\psi=q^{-3/4+ir/3}(\frac{q_{11}}{q_{22}})^{it/2}
(-(1+x^2)f''-2xf'+\frac{t^2}{1+x^2}f).\label{ipd}
\end{eqnarray}
Eigenvalue equation $C_2\psi=\lambda \psi$ gives
\begin{eqnarray}
-(1+x^2)f''-2xf'+(-\lambda+\frac{t^2}{1+x^2})f=0.\label{ipd}
\end{eqnarray}

Compare this with the Legendre equation for a complex argument $z$,
\begin{eqnarray}\label{lc}
(1-z^2)\frac{d^2f}{d z^2}-2z\frac{df}{dz }+(\nu (\nu+1)-\frac{\mu^2}{1-z^2})f=0.
\label{lc}
\end{eqnarray}
It has two linearly independent solutions $\it P_{\nu}^{\mu}(z), \it Q_{\nu}^{\mu}(z)$ over the entire complex plane with appropriately chosen cuts
for any real or complex values of $\nu,\mu$. 
Therefore our eigenfunctions are linear combinations of  $\it P_{\nu}^{it}(iy), \it Q_{\nu}^{it}(iy)$ with $\nu (\nu+1)=-\lambda, ~\mu=it$. 
As discussed above, the relevant values are
$\nu=-\frac{1}{2}+is,~s\in (0,\infty);
~\lambda \in (\frac{1}{4},\infty)$. These functions have appeared earlier. For a recent discussion see Ref.~\cite{d}. 

Legendre equation Eqn.(\ref{lc}) has regular singular points at $z=-1,1,\infty$. As our case is on the imaginary axis $z=iy$, it avoids the singular points at $z=-1,1$. 
Characteristic exponent of these two solutions  at $z=\infty$ are respectively $-\nu,\nu+1$. For our case this gives the asymptotic behaviour $|y|^{-1/2\pm is}$. Thus both linearly independent solutions are equally dominant and behave like the positive and negetive Fourier modes. 
\begin{eqnarray}
\psi(\{q_{ab}\})_{rst}=e^{(-3/4+ir/3)lnq}(\frac{q_{11}}{q_{22}})^{it/2}\it P_{-\frac{1}{2}+is}^{it}(i\frac{q_{12}}{\sqrt q})\label{ipd}
\end{eqnarray}
For $GL(2,R)$ we get the inner product
\begin{eqnarray}
<\phi, \psi>=\int \frac {1}{q^{3/2}}\prod_{ab}~dq_{ab}~ \phi^*(\{g_{cd}\}) \psi(\{q_{cd}\}).
\label{ip2}
\end{eqnarray}
Note that we are interested in only positive definite metrics. Sylvester's criterion says, the necessary and sufficient conditions for a (symmetric) real matrix to be positive definite is that all principal minors be nonnegetive. For us this means
\begin{eqnarray} 
q_{11}\ge 0,q \ge 0.\label{ipd} 
\end{eqnarray}
As a consequence we have
\begin{eqnarray} 
q \ge 0,
q_{11},q_{22}\ge 0,0\le \frac{q_{11}}{q_{22}} \le \infty, -\infty \le \frac{q_{12}}{\sqrt{q}} \le \infty.\label{ipd}
\end{eqnarray}
Changing the variables to $q,q_{11}/q_{22},q_{12}/\sqrt{q}$, we get the invariant measure to be,
\begin{eqnarray}
\int_{-\infty}^{\infty}d(ln~q)  \int_{-\infty}^{\infty} d(\frac{q_{11}}{q_{22}}) \int_{-\infty}^{\infty} d(\frac{q_{12}}{\sqrt{q}})\label{ipd}
\end{eqnarray}
  
There are other well known choices for the basis than are used here. We have isomorphisms of groups $SL(2,R), SU(1,1), SP(1,R)$. For each there is a natural choice of basis. The coset space $SL(2,R)/SO(2)$ can be identified with the Poincare upper half plane and basis is available in Cartesian  and also geodesic coordinate systems. On the other hand the choice of unit disk $\mathbb D$ in the complex plane is useful in many contexts and extensively used.

\section{Noncommutative harmonic analysis on $SL(d,R)/SO(d)$}\label{ha}
We have pointed out relevance of the group $GL(d,R)$ for Einstein gravity. This has a deeper significance.  
At any point of space, the space of metrics relevant to us is the convex cone of  $d\times d$ positive definite real symmetric matrices, $\mathcal Q=\{\{q_{ab}\},q>0\}$. The group $GL(d,R)$ acts on this as Eqn.\ref{gtq}, $G q_{ab}G^{-1}=(g q\tilde g)_{ab}$. The stabilizer group at any point of $\mathcal Q$ is $O(d)$, as seen by the action on the identity matrix which correponds to the  Euclidean metric. Hence $\mathcal Q$ can be identified with the coset space $GL(d,R)/O(d)$. It is a Riemannian symmetric space of noncompact type. 

Consider the subspace $\mathcal S$ of $\mathcal Q$ consisting positive definite real symmetric  matrices of unit determinant. In the same way  as argued above, we get $\mathcal S \simeq SL(d,R)/SO(d)$, also a Riemannian symmetric space of noncompact type.

There is extensive study over decades of what is termed noncommutative harmonic analysis on these spaces. We summarize some of the results relevant for us. Most of the material is to be found in the treatises of Helgason \cite{h1,h2} and 
Diudonn\'{e} \cite{d}. Both $\mathcal Q, \mathcal S$ are special because $(GL(d,R),O(d))$ and $(SL(d,R),SO(d))$ are Gelfand pairs involving the maximal compact subgroup. This leads to very special properties of harmonic analysis. Here we address the space $\mathcal S$. ($\mathcal Q$ can also be directly handled as a reductive homogeneous space using Langland techniques.) For $g \in SL(d,R)$ define  Cartan involution  $\theta: \theta (g)= \tilde g^{-1} $. The manifold of its fixed points is $SO(d)$, the maximal compact subgroup of $SL(d,R)$.

On $G=SL(d,R)$ we have (left and right) $SL(d,R)$ invariant Haar measure $dg$  obtained from the Killing form  $<X,Y>=Tr(ad_X ad_Y)=2d~Tr(XY)$ where $X, Y$ are $d\times d$ matrices representing the Lie algebra elements. $\mathcal S$  inherits $SL(d,R)$ invariant measure $ds$, obtained by keeping only the Lie algebra elements that are invariant under Cartan involution, $ X= \tilde X$. Consider the Hilbert space $L^2(\mathcal S,ds)$ of complex valued functions on coset $\mathcal S$ that are square integrable wrt this measure. Any element $f\in L^2(\mathcal S,ds)$ can be identified with the subspace of functions $f(g) \in L^2(SL(d,R),dg)$  satisfying $f(gO)=f(g)$ for all $O \in SO(d)$. 

Define (left regular) representation of $G=SL(d,R)$ on $L^2(\mathcal S)$ by 
\begin{eqnarray} 
(L(g)f)(s)=f(g^{-1}s).\label{lr}
\end{eqnarray}
This representation is unitary, a consequence of invariance of the Haar measure.

Given two functions $f_1(g), f_2(g) \in L^1(G,dg)$
define their convolution as done in Fourier transform theory: 
\begin{eqnarray} \label{c}
(f_1 \star f_2)(g)=\int f_1(gg_1^{-1})f_2(g_1)dg_1.
\end{eqnarray}
This will also be in $L^1(G,dg)$. Therefore $L^1(G,dg)$ is a Banach algebra. It is not commutative if $G$ is not.
However consider functions $f(g) \in L^1(G,dg)$ which are invariant under both left and right actions by $SO(d)$: $f(gO_1)=f(O_2g)=f(g)$ for any $O_1,O_2 \in SO(d)$. These are functions on the double coset $SO(d) \backslash G/SO(d)$ and form a closed subalgebra of the Banach algebra $L^1(G,dg)$. {\it Because $SO(d)$ is the fixed point of an involution}, this algebra is commutative. As a consequence, the algebra $D(\mathcal S)$ of bounded linear operators on $L^2(\mathcal S)$ which are invariant under the full group $G$ is commutative. Consider simultaneous eigenfunctions $\phi_{\lambda}(g)$ of this commuting set of operators on the double coset $SO(d) \backslash G/SO(d)$. For any $D \in \mathbf D(\mathcal S)$,
\begin{eqnarray}
D\phi_{\lambda}(g)=\chi_{\lambda}(D)\phi_{\lambda}(g).
\label{sp}
\end{eqnarray}
Any such $\phi_{\lambda}(g)$ is called a {\it spherical function of positive type} for reasons explained below. The corresponding eigenvalue $\chi_{\lambda}(D)$ is called a character. In our case, for any given character the spherical function of positive type is unique, i.e. {\it the vector space of eigenfunctions is one dimensional}.

Now consider the subspace of joint eigenfunctions in the larger space $\mathcal S$ corresponding to a character $\chi_{\lambda}(D)$ 
\begin{eqnarray}
Df_{\lambda}(g)=\chi_{\lambda}(D) f_{\lambda}(g).
\label{ei}
\end{eqnarray}
These functions span an irreducible unitary representation,
\begin{eqnarray} 
(\pi_\lambda(g)f_{\lambda})(g_1)=f_{\lambda}(g^{-1}g_1).\label{gf}
\end{eqnarray}
This gives a complete decomposition of $L^2(\mathcal S)$ into irreducible unitary representations of $SL(d,R)$:  $L^2(\mathcal S)=\int_\lambda^\oplus \mathcal {H}_\lambda (\mathcal S)$.

In case of a symmetric space $G/K$ with compact group $G$ and a maximal subgroup $K$ obtained by a Cartan involution, we have the
Cartan's theorem
\begin{eqnarray} 
f=\sum_{\delta \in \hat {G}_H}d(\delta)f\star\phi_\delta,~ f\in C_c^{\infty}(G/K),\label{ct} 
\end{eqnarray}
where $\phi_\delta$ is the spherical function and $d(\delta)$ the degree of an irreducible unitary representation $\delta$. Not all irreducible unitary representations of $G$ appear in the decomposition above. Only the subset $G_H$ for which there is a vector $v_0$ invariant under the subgroup $K$ appear.

For our noncompact case we have very similar result:
\begin{eqnarray} 
f(g)=\int_{\mathfrak a^*} f\star \phi_\lambda(g) |{\bf c}(\lambda)|^{-2} d \lambda, ~ f\in D({\mathcal S}).\label{ft}
\end{eqnarray}
$\bf {c}(\lambda)$ is the Harish-Chandra c-function and $d\lambda$ is a Euclidean measure elaborated below. Only the principal series of irreducible unitary representations of $G$ appear in this decomposition. This explains why only one class of representations of Bargmann appeared in explicit calculations in Sec.\ref{2} with $SL(2,R)$.

It is possible to introduce a measurable family of complete orthonormal system $v_0(\lambda),v_1(\lambda),v_2(\lambda),\cdots$in $\mathcal {H}_\lambda$ 
such that $v_0(\lambda)$ is invariant under $SO(d)$ action, i.e. $\pi_\lambda(O)v_0(\lambda)=0, \forall O \in SO(d)$, and the subspace spanned by the rest of the orthonormal system does not have any non-zero elements invariant under all $\pi_\lambda(O)$. We get
\begin{eqnarray} 
\phi_\lambda(g)=<\pi_\lambda(gO)v_0(\lambda)|v_0(\lambda)>.\label{sp1}
\end{eqnarray}
Define 
\begin{eqnarray} 
\hat F(\lambda)=\int f(g)\pi_\lambda(g) dg,~ f\in L^1(\mathcal S)\cap L^2(\mathcal S).\label{ft1}
\end{eqnarray}
This is a bounded operator on $\mathcal H_\lambda$ for almost every $\lambda$. Moreover
\begin{eqnarray}
<\hat F(\lambda)v_m(\lambda)|v_n(\lambda)>=0 ~if~ n \ne 0, \forall m.\label{ft2}
\end{eqnarray}
This allows us to define the analogue of Fourier modes, 
\begin{eqnarray}
\tilde f_m(\lambda)=<\hat F(\lambda)v_m(\lambda)|v_0(\lambda)>.\label{fm}
\end{eqnarray}
From unitarity of the representation we get 
\begin{eqnarray}
\int_{\mathcal S} |f(s)|^2 ds=\int_{\mathfrak a^*}\sum_m|\tilde f_m(\lambda)|^2|{\bf c}(\lambda)|^{-2}d\lambda.\label{pt}
\end{eqnarray}
This is the Plancherel theorem giving an isometry  $f(s) \leftrightarrow \tilde{f}_m(\lambda)$.

We now give technical details of the summary stated above.

Any element  $g \in SL(d,R)$ can be uniquely and continuously decomposed as $g=O(g)A(g)N(g),O(g)\in SO(d), A(g) \in A, N(g) \in N$. Here $A$ is the subgroup of $d\times d$ diagonal matrices with positive diagonal entries $A_i$, product of the diagonal elements is 1: $\Pi_i A_i=1$. $N$ is the nilpotent subgroup of all real upper triangular matrices with diagonal entries $1$. This is the Iwasawa decomposition giving an analytic map of $SL(d,R)$ into $SO(d,R),A,N$. The mapping 
$N\times A \rightarrow \mathcal S: (n,a)\rightarrow na\tilde{(na)}$ is a diffeomorphism. 

Corresponding to the Iwasawa decomposition we have decomposition of the Lie algebra, $\mathfrak g= \mathfrak l \oplus \mathfrak a \oplus \mathfrak p$. Here $\mathfrak l$ is the Lie algebra of $SO(d)$. $\mathfrak a=\{diag(a_1,a_2,\cdots, a_d);a_j \in R,\sum_{j}a_j=0 \}$ is the maximal abelian subalgebra. The rank $r$ of the group is given by the dimension $(d-1)$ of $\mathfrak a$. The (real) dual of $\mathfrak a$ is denoted by $\mathfrak a^*$ and the complexification of this dual by $\mathfrak a_C^*$. 

For any $\lambda \in \mathfrak a^*$ define $g_\lambda =\{X \in g: [H,X]=\lambda (H)X, \forall H\in \mathfrak a\}$, i.e. $g_\lambda$ is the corresponding  simultaneous eigenspace. If $\lambda \ne 0$ and $g_\lambda \ne 0$ such a $\lambda$ is called a restricted root and any $X \in g_\lambda$ is called a restricted root vector. The set of all restricted roots is denoted by $\Sigma$. 

$H\in \mathfrak a$ is called regular if $\lambda (H) \ne 0, \forall \lambda \in \Sigma$, otherwise it is called singular. Denote the set of regular points by $\mathfrak a' \in \mathfrak a$. For each root consider the hyperplane $\lambda (H)=0$. These hyperplanes divide the space into finitely many connected components called Weyl chambers. Fix a Weyl chamber $\mathfrak a^+$ and call a restricted root $\lambda$ positive if it has positive values on $\mathfrak a^+$. The set of all positive restricted roots is denoted by $\Sigma^+ $.

A root  $\lambda \in \Sigma^+ $ is called simple if it is not a sum of two positive roots. The number of simple roots is equal to the rank $r$ of the group. Denote them by $\lambda_1,\lambda_2, \cdots, \lambda_r$ and the corresponding dual basis by $H_1, H_2, \cdots H_r$. Therefore $\mathfrak a^+=\{H\in \mathfrak a,\lambda_1(H)>0,\lambda_2(H)>0,\cdots \lambda_r(H)>0\}$. Let $\mathfrak a^+$ be lexicographically ordered wrt this basis.

Define $\epsilon_j \in  \mathfrak a^*$ by $\epsilon_j(diag(a_1,a_2,\cdots, a_d))=a_j$. For $SL(d,R)$ the set $\Sigma$ of restricted  roots of $(\mathfrak g,\mathfrak a)$ is 
$\Sigma=\{\alpha_{ij}=\epsilon_{i}-\epsilon_{j},1\le i \ne j \le d \}$. All root multiplicities $m_\alpha$ are $1$. The set of positive roots is $\Sigma^+=\{\alpha_{ij};i < j\}$. The $(d-1)$ simple roots are $(\epsilon_{1}-\epsilon_{2}),(\epsilon_{2}-\epsilon_{3}),\cdots, (\epsilon_{(d-1)}-\epsilon_{d})$. The Weyl group $W$ of $\Sigma$ is the group of permutations $S_d$ acting on $\epsilon_1,\epsilon_2,\cdots,\epsilon_d$. The Lie algebra of the nilpotent subgroup $N$ is simply $\mathfrak n=\Sigma_{\lambda \in \Sigma^+}\mathfrak g_\lambda$.

Let $A=e^{\mathfrak a}$, $A^+=e^{\mathfrak a^+}$
and $\bar {A^+}$ the closure of $A^+$ in G. Then we have the Cartan decomposition $G=O\bar {A^+}O$, i.e. $g=O_1(g)A^+(g)O_2(g)$ uniquely, where $A^+(g) \in \bar {A^+}$. We write $A^+(g)=exp~{a^+(g)}$ where $a^+(g) \in {\bar a^+}$.

We write the Iwasawa decomposition as $g=O(g)exp(H(g))N(g)$ where the unique $H(g) \in \mathfrak a$ is called the H-function and plays an important role. After all the machinary described above, we can present Harish-Chandra's explicit formula for the spherical functions of positive type.  They are {\it completely} given by 
\begin{eqnarray} 
\phi_{\lambda}(g)=\int_O e^{(i\lambda-\rho)H(gO)}dO\label{sph},
\end{eqnarray}
for all $\lambda \in a^*$. Here $\rho=1/2 \sum_{\lambda \in \Sigma^+} \lambda$.
Also all such $\phi_{\lambda}$ are distinct unless  the corresponding $\lambda$ are related by a Weyl transformation: $\phi_{\lambda}=\phi_{\mu}$ if $\lambda =W \mu$.
$\lambda$ is called the spectral parameter. Thus $\{\lambda ~mod~ W, \lambda\in \mathfrak a^*\}$ provides a unique labelling of all irreducible unitary representations of $SL(d,R)$ appearing in harmonic analysis on $\mathcal S$.

$\phi_{\lambda}$ has many important propeties. i. 
$\phi_{\lambda}(g)=\phi_{-\lambda}(g^{-1})=\bar\phi_{-\lambda}(g)$.
ii. Since $\phi_{\lambda}(O_1gO_2)=\phi_{\lambda}(g)$, it is a function on the double coset. iii. Also $\int \phi_{\lambda}(g_1Og_2)dO=\phi_{\lambda}(g_1)\phi_{\lambda}(g_2)$ where $dO$ is an integration over all $O \in SO(d)$. iv. $|\phi_{\lambda}(g)|\le \phi_{\lambda}(e)=1$. v. If $f$ is a square integrable function in the double coset then  $f*\phi_{\lambda}=\lambda_f \phi_{\lambda}$ with $\lambda_f=\int_G  f(g)\bar \phi_{\lambda}(g)dg$. vi. An extremely important property is that $\phi_{\lambda}(g)$ \textit {is a function of positive type}: $\sum_{ij}z_i\bar z_j\phi_{\lambda}(g_i^{-1}g_j) \ge 0$ for any set $\{g_i \in G\}$ and $\{z_i \in C\}$. vii. It is an analytic function. 

Every positive definite matrix can be diagonalized by an orthogonal transformation. Therefore the {\it polar decomposition} $O\times A \rightarrow \mathcal S, (o,a)\rightarrow oa^2\tilde o$ is an onto mapping. The generic fiber is isomorphic to a finite group, the Weyl group $W$. This acts as the permutation group $S_d$, permuting the diagonal elements $A_i$. Therefore we get the identification $C^\infty(\mathcal S)^O\simeq C^\infty(A)^W$. Here the lhs is the space of smooth $SO(d)$ invariant functions on $\mathcal S$ and rhs is smooth functions of 
$\{A_i\}$ invariant under permutations. Consider the action of any  $D \in \mathbf D(\mathcal S)$ on a function in the double coset $SO(d,R) \backslash G/SO(d,R)$. The corresponding operator $D_A$ is called the radial operator. This operator is automatically inavariant under the Weyl group $W$, $ D_A\in D_W(A)$. 

We give technical definition of the $\it radial~ part$ $R_N(D)$ of any  operator $D \in \mathbf D(\mathcal S)$ wrt the subgroup $N$. Let $\overline f$ denote the restriction of a function $f \in \mathcal S$ to $A$. Then for any operator $D \in \mathbf D(\mathcal S)$ there is a unique operator $ R_N(D)\in D_W(A)$ called the radial part of $D$ wrt $N$ such that 
\begin{eqnarray}
\overline {(Du)}=R_N(D)\overline u,\label{d}
\end{eqnarray}
for each locally invariant function $u$.
The explicit form is 
\begin{eqnarray}
R_N(D)=e^\rho D_A\circ e^{-\rho}-|\rho|^2 .\label{d1}
\end{eqnarray}
Here the norm $|\rho|$ is defined as follows. The Killing form defines a metric on $\mathfrak a$ and by duality a inner product on $\mathfrak a^*$. As it is positive definite it gives us the norm $|\lambda|=\sqrt {<\lambda|\lambda>}$.
The mapping 
\begin{eqnarray}
\Gamma:D \rightarrow e^{-\rho}R_N(D) \circ e^{\rho}.\label{d2}
\end{eqnarray}
is an isomorphism of  $\mathbf D(\mathcal S)$ onto $\mathbf D_W(A)$, the Harish-Chandra isomorphism.  

The foremost member of $D(\mathcal S)$ is the Laplacian operator $\Delta$ of the Riemannian space which is also the Casimir operator $tr(T^2)$. We have 
$\Gamma (\Delta)=\Delta_A-|\rho|^2$. The entire algebra is generated by the basic set $tr(T^n), n=2,3,\cdots d$. (Note that $tr(T)=0$). Spherical functions of positive type are eigenfunctions of the Laplacian: $\Delta \phi_{\lambda}=-(|\lambda|^2+|\rho|^2)\phi_{\lambda}$.
Explicit calculation in Sec.\ref{2} gave the spectrum of the Laplacian in $d=2$ as $(1/4,\infty)$. In $d=3$ it ranges over $(1/3,\infty)$.

Sekiguchi \cite{seki} gives an explicit formula for the generating function of the basic $(d-1)$ radial operators $\Delta_i, i=2,3,\cdots d$:
\begin{eqnarray}
\Delta(\zeta)=\frac {1}{\delta(H)}\sum_{s\in W}det(s) ~e^{2\rho(sH)}\Sigma_{i=1}^d(\xi+D_{s(i)}+\frac {d+1-2i}{2})
=\zeta^d+\Delta_1\zeta^{d-1}+\Delta_2\zeta^{d-2}+\cdots +\Delta_d.\label{seki}
\end{eqnarray}
Here
\begin{eqnarray}
D_i=\frac {\partial}{\partial t_i},~\delta(H)=\Pi_{i<j}(e^{t_i-t_j}-e^{t_j-t_i}),~
\rho(H)=\frac {1}{2}\Sigma_{i<j}(t_i-t_j)\\
s(H)=(t_{s(1)},t_{s(2)},\cdots ,t_{s(d)}),\label{seki2}
\end{eqnarray}
for $H=(t_1,t_2,\cdots,t_d)\in \mathfrak a$, and $s$ label the elements of the Weyl group W. In particular the radial part of the Casimir operator is
\begin{eqnarray}
\Delta_2=\sum_{i<j}(D_iD_j-\frac {1}{2}coth(t_i-t_j)(D_i-D_j))-\frac {1}{2}|\rho|^2\label{seki3}.
\end{eqnarray}
$\Delta(\zeta)$ is a family of commuting operators. Consider the eigenvalue equation 
\begin{eqnarray}
\Delta (\zeta) u=\Pi_1^d (\zeta+\lambda_i) u.\label{seki4}
\end{eqnarray}
for any real values of $(\lambda_1,\lambda_2,\cdots,\lambda_d), \sum_1^d \lambda_i=0$
Explicit solution are given in . They are generalizations of Gegenbauer polynomials.

We can define radial operator wrt the Cartan decomposition as we did with the Iwasawa decomposition. Now the Laplacian has the form 
\begin{eqnarray}
R_K(\Delta)=\Delta_A+\sum_{\alpha \in \Sigma^+} coth~\alpha~ H_{\alpha}
\label{l3}
\end{eqnarray}
This is Schr\"{o}dinger operator for Calagero-Moser model.

Harish-Chandra has derived a formula for the $\bf c$-function. Gindikin and Karpelevik \cite{gk} have obtained an explicit product formula for $SL(d,R)$.
Similar product formula is available \cite{bo} also for the spherical functions.

\section{Discussion}\label{dis}
We rewrite the constraints Eqn.~\ref{hc},\ref{mc} using $GL(3,R)$ generators.
The Hamiltonian constraints of ADM formalism takes the form
\begin{eqnarray}
(\mathbf{T}_a^b\mathbf{T}_b^a-\frac{1}{2}\mathbb T^2)(X)-\frac{q(X)}{4\kappa^2}R^{(3)}(X)=0, \label{hc1}
\end{eqnarray}
and the momentum  constraints,
\begin{eqnarray}
\partial_{b}\mathbf{T}_a^b(X) +\mathbf{P}_a(X)=0.\label{mc1}
\end{eqnarray}
$\mathbf{P}_a(X)$ is the generator of local translations in argument $X$. Its role is a local translation of the space coordinate $X$, while the first term in Eqn.~\ref{mc1} handles the tensorial properties under general coordinate transformations. There are important ordering issues when these constraints are derived \cite{I} from the naive functional integral of Einstein gravity. This is handled in detail in  Ref.\cite{III}, where we also include the effects of an external energy-momentum source. 

It is now clear that it is natural to use noncommutative harmonic analysis on $GL(d,R)/O(d)$ as presented in this paper. We  obtain the general solution of the Hamiltonian constraints Eqn.\ref{hc1} in  Ref.\cite{III}. We also obtain the general  solution of the momentum  constraints Eqn.\ref{mc1} in case of space dimension $d=2$. 

In this paper we have highlighted that the  constraints of the Hamiltonian  formalism of Einstein gravity are to be interpreted through local $GL(d,R)$ Lie group. Then the machinary of noncommutative harmonic analysis  developed over many decades can be used for detailed calculations of quantum gravity. This provides a powerful tool to address the many mysteries of quantum gravity.


\begin{thebibliography}{1} \bibliographystyle{plain}
\bibitem{adm} R. Arnowitt, S. Deser and C.W. Misner, {The Dynamics of General Relativity. Gravitation: An introduction to current research. Witten L., editor, Wiley, N.Y.}, Reprinted as arXiv gr-qc/0405109.
\bibitem{I} H S Sharatchandra, "Quantization of  Einstein Gravity: Extraction of Hilbert Space and Constraints",  arXiv 1806.04097.
\bibitem{nist}F. W. J. Olver, A. B. Olde Daalhuis, D. W. Lozier, B. I. Schneider, R. F. Boisvert, C. W. Clark, B. R. Miller, B. V. Saunders, H. S. Cohl, and M. A. McClain, eds.,\emph{NIST Digital Library of Mathematical Functions}, http://dlmf.nist.gov/, Release 1.0.25 of 2019-12-15. 
\bibitem{du}T.M. Dunster, "Conical functions of purely imaginary order and argument", Proc. Roy. Soc. Edinburgh Sect. A 143(A) pp 929-955.	
\bibitem{h1}S Helgason, Geometric Analysis on Symmetric Spaces, V 39, Mathematical Surveys and Monographs, American Mathematical Soceity, Providence RI, 2nd Edition 2008.	
\bibitem{h2} S Helgason, Groups and Geometric Analysis. Integral Geometry, Invariant Differential Operators, and Spherical Functions. Pure and Applied Mathematics, 113, academic Press, inc., Orlando, FL, 1984.
\bibitem{d}J Dieudonn\'e, Special functions and Linear representations of Lie Groups, Regional Conference Series in Mathematics, no. 42, AMS First Ed. 1980.
\bibitem{seki} J Sekiguchi, "Zonal Spherical Functions on some Symmetric Spaces", Pub. RIMS Kyoto univ. {\bf 12} Suppl. (1977), 455-464.
\bibitem{gk} S G Glindikin, S I Karpelevi\u {c}, Dokl. Akad. Nauk.SSSR {\bf 145} (1962) 252-255.
\bibitem{bo}U N Bassey, O O Oyadare, "Helgason-Schiman Formula for Semisimple Lie groups of Arbitrary Rank", J Generalized Lie Theory Appl. (2014) 9.1.DOI 10.4172/1736-4337.1000216.
\bibitem{III} H.S.Sharatchandra, "Solution of quantum ADM constraints with external energy momentum tensor using local gl(d,R) Lie algebra", in preparation.
%
\end{thebibliography}
\end{document}